\documentclass[fleqn,usenatbib]{mnras}

\usepackage{newtxtext,newtxmath}

\usepackage[T1]{fontenc}

\DeclareRobustCommand{\VAN}[3]{#2}
\let\VANthebibliography\thebibliography
\def\thebibliography{\DeclareRobustCommand{\VAN}[3]{##3}\VANthebibliography}

\usepackage{graphicx}	
\usepackage{amsmath}	
\usepackage{mathtools}
\usepackage{lipsum} 
\usepackage{pifont}

\usepackage{multicol}
\usepackage{comment}

\title[NS-XRBs on the radio:X-ray plane]{The 2019 outburst of AMXP SAX J1808.4--3658 and radio follow up of MAXI J0911--655 and XTE J1701--462}

\author[K. V. S. Gasealahwe et al.]{K. V. S. Gasealahwe,$^{1,2}$\thanks{E-mail: kelebogile@saao.ac.za}
I. M. Monageng,$^{1,2}$
R. P. Fender,$^{1,3}$
P. A. Woudt,$^{1}$
S. E. Motta,$^{3,4}$
J. van den Eijnden,$^{3}$
\newauthor D. R. A. Williams, $^{5}$
I. Heywood,$^{3,6,7}$
S. Bloemen,$^{8}$
P. J. Groot,$^{1,2,8}$
P. Vreeswijk,$^{8}$
V. McBride,$^{1,2,9}$
M. Klein-Wolt,$^{8}$
\newauthor E. K\"{o}rding,$^{8}$
R. Le Poole,$^{10}$
D. Pieterse$^{8}$
and S. de Wet$^{1}$
\\\\
$^{1}$Department of Astronomy, University of Cape Town, Private Bag X3, 7701 Rondebosch, South Africa\\
$^{2}$South African Astronomical Observatory, P.O. Box 9, 7935
Observatory, South Africa \\
$^{3}$Department of Physics, University of Oxford, Denys Wilkinson Building, Keble Road, Oxford OX1 3RH, UK \\
$^{4}$Istituto Nazionale di Astrofisica, Osservatorio Astronomico di Brera, via E. Bianchi 46, 23807 Merate (LC), Italy \\
$^{5}$Jodrell Bank Centre for Astrophysics, School of Physics and Astronomy, The University of Manchester, Manchester, M13 9PL, UK \\
$^{6}$Department of Physics and Electronics, Rhodes University, PO Box 94, Makhanda 6140, South Africa \\
$^{7}$South African Radio Astronomy Observatory, 2 Fir Street, Black River Park, Observatory 7925, South Africa\ \\
$^{8}$Department of Astrophysics/IMAPP, Radboud University, P.O. 9010, 6500 GL, Nijmegen, The Netherlands \\
$^{9}$IAU Office of Astronomy for Development, P.O. Box 9, 7935
Observatory, South Africa \\
$^{10}$Leiden Observatory, Leiden University, P.O. Box 9513, NL-2300 RA Leiden, The Netherlands \\
}

\date{Accepted XXX. Received YYY; in original form ZZZ}

\pubyear{2022}

\begin{document}
\label{firstpage}
\pagerange{\pageref{firstpage}--\pageref{lastpage}}
\maketitle

\begin{abstract}
 We present radio coverage of the 2019 outburst of the accreting millisecond X-ray pulsar SAX J1808.4--3658, obtained with MeerKAT. We compare these data to contemporaneous X-ray and optical measurements in order to investigate the coupling between accretion and jet formation in this system, while the optical lightcurve provides greater detail of the outburst. The reflaring activity following the main outburst peak was associated with a radio re-brightening, indicating a strengthening of the jet in this phase of the outburst. We place quasi-simultaneous radio and X-ray measurements on the global radio:X-ray plane for X-ray binaries, and show they reside in the same region of luminosity space as previous outburst measurements, but significantly refine the correlation for this source. We also present upper limits on the radio emission from the accreting millisecond X-ray pulsar MAXI J0911--655 and the transitional Z/Atoll-type transient XTE J1701--462. In the latter source we also confirm that nearby large-scale structures reported in previous radio observations of the source are persistent over a period of $\sim 15$ years, and so are almost certainly background radio galaxies and not associated with the X-ray transient.
\end{abstract}




\section{Introduction}
Neutron Star (NS) X-ray binaries (XRBs) are binary systems comprising a NS that accretes matter from a companion star. A subgroup of these are systems with low-mass (M$_{\rm companion}$ $<$ 1 M$_\odot$) companions, and are thought to be considerably older than systems with a higher mass (M$_{\rm companion} \gtrsim8$ M$_\odot$; \citealt{1978ApJ...221..645J}). In a subset of the low-mass systems, the NS is observed to spin at periods significantly smaller than a second, either through the detection of pulsations in their X-ray flux or oscillations during thermonuclear Type I X-ray bursts on their surface \citep{2021ASSL..461..143P}. In accreting millisecond X-ray pulsars (AMXPs), of which SAX J1808.4--3658 \citep{1998ApJ...507L..63W} is the prototype, the accretion and millisecond combination may be explained by the scenario of recycled pulsars. In this scenario the pulsar originates as a `normal' pulsar that eventually spins down on a timescale of t $\approx 10^6$\,yrs, possibly in combination with magnetic field decay. The pulsar is spun up again through an accretion phase and manifests as an AMXP \citep{1982Natur.300..728A,1982Natur.300..615B,2010NewAR..54...93S} which emits pulsed X-rays with a period of 1-10\,ms. Some pulsars are estimated to have a relatively low magnetic field strength of $\sim$ 10$^8$\,G. Two other empirical classifications of NS-XRBs at high accretion rate and low magnetic field ($\leq 10^9$\,G) systems are known: the Atoll and Z sources. The former may also exhibit pulsations. These two classes are distinctly classified according to the ratio of their X-ray soft and hard energy bands often referred to as X-ray colours, traced over different time-scales \citep{1989A&A...225...79H}. For the Z sources, the hard and soft colours plotted against each other reveal a 'Z-shape' pattern traversed on time-scales of hours to days, while Atoll sources show a 'C-shape' pattern generally accompanied by an 'island' at large hard colours on time-scales ranging from weeks to months \citep{2002ApJ...568L..35M}. Some sources display a blend of properties of the two classes as well as properties that enter in neither (e.g. Cir X-1; \citealt{2011MNRAS.414.3551M}). Most dramatically the neutron star transient XTE J1701--462 demonstrated a clear transition from Z-like to Atoll-like properties as its luminosity dropped resulting from a high mass transfer rate ($\dot{m}$) to a lower $\dot{m}$ during the outburst decay. This allowed to connect the two classes to accretion rate \citep{2007ApJ...656..420H}. This blurring of some Atoll and Z sources coupled with properties of AMXPs makes the distinction of the different classes of NS-XRBs poorly understood.
\\\\
Interesting in their own right, accreting low-magnetic field neutron stars in the Z/Atoll/AMXP classes play a crucial role as a control sample to the Black Hole (BH) population in the context of the coupling between accretion and jet formation in BHs, which have no surface. While detailed studies of individual sources are important, a key global diagnostic of this jet:outflow (where the radio is from the jets and X-rays are observed from accretion flow) coupling is the use of the radio:X-ray luminosity plane \citep{2003MNRAS.344...60G}. One key result from such global studies is that the sample of black holes are typically more radio-loud (exhibiting brighter radio luminosities) than the populations of neutron stars for a given X-ray luminosity \citep{2001MNRAS.324..923F,2006MNRAS.366...79M}. Another key result is the slope of the radio:X-ray correlation. In BH systems in the hard X-ray state (or: at relatively low luminosities) accretion is believed to be inefficient, as a large fraction of the gravitational potential energy is not radiated, but lost across the BH event horizon \citep{2012Sci...337..540F}. In contrast, in NS systems accretion may be efficient as all the potential energy of the accreting matter is released at the NS surface. This may result in a steeper correlation (see e.g. discussion in \citealt{2011IAUS..275..255C}). Multiple studies of NS systems, however, suggest that the slope of the correlation for some NS-XRBs is inconsistent with this assumption \citep{2006MNRAS.366...79M,2009MNRAS.400.2111T,2018MNRAS.478L.132G,2015ApJ...809...13D}. The correlation slope of some NSs being similar to that of BHs is surprising because the physical properties are different between the compact objects. In order to develop a better understanding of the radio:X-ray correlation for NS-XRBs, a quantitative analysis of luminosity strength is required of these systems. In this study we do not consider high-magnetic field accreting neutron stars, which are less radio luminous and may well have a quite different flow geometry close to the NS, but we direct the interested reader to \cite{2021MNRAS.507.3899V} for details.
\\\\
SAX J1808.4–3658 (SAX J1808 hereafter), is an X-ray binary with an X-ray pulsation at 401 Hz. SAX J1808 was discovered in 1998 by the \textit{Rossi X-ray Timing Explorer (RXTE)} \citep{1996ApJ...469L..33L}, and was the first confirmed AMXP \citep{1998ApJ...507L..63W}. SAX J1808 displays Type I thermonuclear X-ray bursts based on which a distance of 3.5 kpc was estimated \citep{1998A&A...331L..25I, 2006ApJ...652..559G}. An orbital period of $\approx$ $2$~hr \citep{1998Natur.394..346C} is known for the source. SAX J1808 is reported to go into semi-regular outbursts with recurrence times between 2 and 4 years. Previous outbursts from SAX J1808 were recorded in 1999, 2002, 2005, 2015 \citep{2017MNRAS.470..324T}, and the latest one occurred in 2019. In this paper we report on the 2019 outburst of SAX J1808 as observed with MeerKAT (in radio \citealt{2019ATel13026....1W}). SAX J1808 was observed with the X-ray telescope (XRT) onboard the Neil Gehrels Swift telescope \citep[\textit{Swift};][]{2005SSRv..120..165B} and the \textit{Neutron Star Interior Composition ExploreR} (\textit{NICER}; \citealt{gendreau2012}) in the X-rays and additionally show the optical Las Cumbres Observatory LCO data reported in \cite{2020ApJ...905...87B} to trace the main outburst via a multi-wavelength approach, and to measure the radio:X-ray correlation for this outburst of the source. 
\\\\
Additionally, while we compare SAX J1808 to the rest of the AMXP radio and X-ray luminosity  measurements (L$_{\rm R}$L$_{\rm X}$), we also report short MeerKAT observations of the AMXP MAXI J0911--655 (MAXI J0911 hereafter, \citealt{2016ATel.8872....1S,2016ATel.8971....1H}) and transitional Atoll/Z-source XTE J1701--462 (XTE J1701 hereafter, \citealt{2007ApJ...656..420H}). 

\section{Observations}
\subsection{MeerKAT observations}
The 2019 outburst of SAX J1808 was monitored with MeerKAT at 1.28\,GHz, through the ThunderKAT Large Survey Programme \citep{2016mks..confE..13F} and by {\textit{Swift}} as part of an associate observing program. MeerKAT consists of 64 dishes and on average 61 dishes were available per observation. The outburst was monitored nearly weekly for 6 epochs (Table~\ref{tab:lums}) from 2019 July 31 to August 31, for 15\,mins every week. J1939--6342 was observed for 5\,mins at the start of each observation, as the primary calibrator for flux and bandpass calibrations. The secondary calibrator, J1830--3602, was observed for 2\,mins before and after the target in each observation. All observations were performed with a total bandwidth of 860 MHz, divided into 4096 channels of width 209\,kHz. The MeerKAT data were reduced using the semi-automated pipeline, {\textsc{oxkat}}\footnote{for more details see, \url{https://github.com/IanHeywood/oxkat}} \citep{2020ascl.soft09003H}. The first generation process of {\textsc{oxkat}} involves making use of {\textsc{casa}} \citep{2007ASPC..376..127M} to perform averaging, flagging and do cross- and self-calibration. Once the visibility and gain solution plots produced were satisfactory after inspection, the data were then flagged and imaged using the {\textsc{tricolour}} and {\textsc{wsclean}} packages respectively \citep{2014MNRAS.444..606O}. Thereafter, the second generation processing was initiated to do self calibration, plot gain solutions and a second output of images were produced. The flux densities were then determined using {\textsc{pybdsf}} \citep{2015ascl.soft02007M}.
\\\\
MAXI J0911 and XTE J1701 were observed with MeerKAT in the same way as SAX J1808. MAXI J0911 was observed around a phase of hard X-ray emission \citep{2021ATel14767....1N} on the 8 and 26 July 2021 while XTE J1701 was observed on the 26 July 2021 as part of a broader study of NS-XRBs with ThunderKAT. The sources were observed for 15\,min for each observation and the reduction and analysis procedures are the same as those described above for SAX J1808, except the primary and secondary calibrator for MAXI J0911 were J0408-6545 and J0906-6829 respectively, and the secondary calibrator for XTE J1701 was J1744-5144. 

\subsection{X-ray observations}
\begin{figure*}
    \centering
    \includegraphics[scale=0.5]{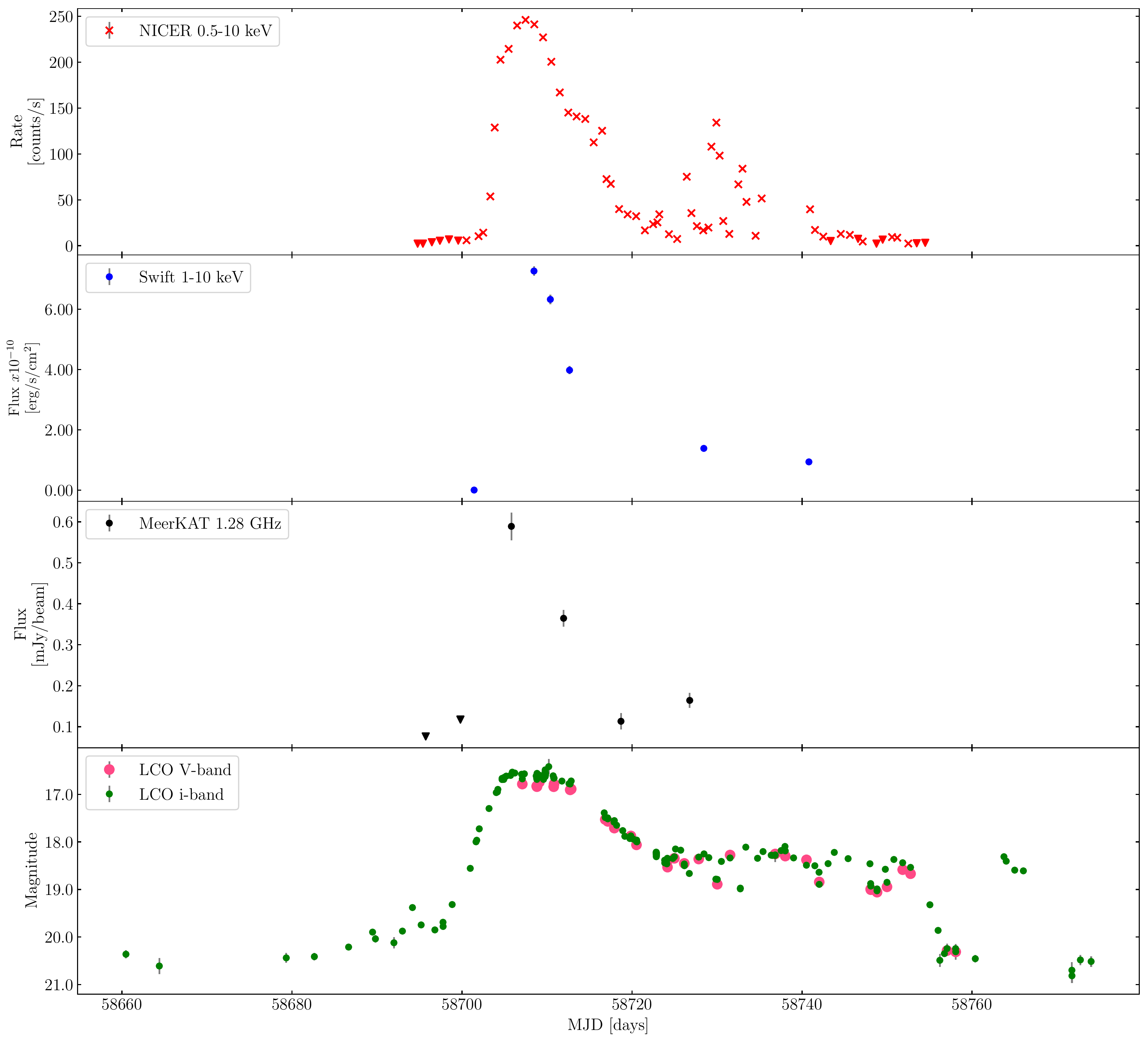}
    \caption{The 2019 outburst of SAX J1808 as seen in the X-ray (top panel: 0.5--10\,keV, second panel: 1--10\,keV), radio (third panel: 1.28 GHz) and optical (bottom panel: $V$ and $i$ band). The upper limits from \textit{NICER} and the non-detections of the first two epochs in the radio are indicated with the red and black upside down triangles, respectively.}
    \label{fig:SAX_flux}
\end{figure*}

\textit{Swift}/XRT provided a total of 21 observations of SAX J1808 of variable duration, of which 10 were between 31 July to 31 August 2019, taken quasi-simultaneously with MeerKAT. 
The overall \textit{Swift}/XRT exposure time was 
formed by a number of snapshots, taken in both Windowed Timing (WT) mode and Photon Counting (PC) mode \citep{2005SSRv..120..165B}, both providing data in the 0.3--10.0~keV energy band. Following the XRT reduction threads \footnote{\url{http://www.swift.ac.uk/analysis/xrt/\#abs}} we extracted spectra from circular regions centred at the source position, with a radius of 30 pixels. When fitting the energy spectra we ignored data below 0.6 keV since spectra can be dominated by strong redistribution effects associated with the WT readout process, as well as trailing charge released from deep charge traps in the CCD on time-scales comparable to the WT readout time. The above might result in additional low energy events that may distort the spectrum. Spectra were then extracted following standard extraction procedures \footnote{See, e.g., \url{https://www.swift.ac.uk/analysis/xrt/spectra.php}}. 
\\\\
Based on the radio observations taken of MAXI J0911 and XTE J1701, we followed up to check if any observations were taken with \textit{Swift} at the time. \textit{Swift}/BAT survey data was found of both sources, which were within a day of quasi-simultaneity of the first epoch of MAXI J0911 radio observations and the single epoch of XTE J1701. 
\\\\
The \textit{NICER} telescope observed the 2019 outburst of SAX J1808 extensively in X-rays, providing near-daily coverage of the outburst between 30 July and 28 September 2019 \citep{2019ApJ...885L...1B}. In order to probe the overall evolution of the outburst, we analysed the cumulative data taken on each day, often combining a number of shorter exposures per MJD. We applied the \textsc{ftool} \textsc{nicerl2} to perform the most recent calibration, after which we extracted the source count rate between $0.5$ and $10$ keV. To calculate the background count rate, we applied the \textit{nicer\_bkg\_estimator} model, based on \textit{RXTE} blank sky background fields \citep{jahoda2006}, in the same energy band. During days where the source count rates does not exceed three times the background, we report $3\sigma$ upper limits. For the remaining observations, we show the light curve in the top panel of Figure~\ref{fig:SAX_flux}. Due to the small background rate compared to the source count rate during the majority of the outburst, for simplicity we plot the source count rates not background-corrected. 
Both the \textit{Swift}/XRT and \textit{NICER} spectra were fitted with a model in the form ($\rm{tbfeo*[continuum]}$), where tbfeo is an ISM model and \textit{[continuum]} was either a powerlaw (\textsc{power} in \textsc{XPEC}), a disk-blackbody component (\textsc{diskbb}), or a combination of disk-blackbody and a blackbody component (\textsc{diskbb} + \textsc{bbodyrad}). In modelling the interstellar absorption, we assumed abundances from \cite{2000ApJ...542..914W} and cross-sections from \cite{1996ApJ...465..487V}. The \textit{Swift}/XRT and \textit{NICER} spectra were fitted in the 0.6--10keV and 0.2--10keV energy range, respectively. In both cases, we fixed the equivalent Hydrogen column density to n$_{\rm{H}}$ $= 0.117\times10^{22} \rm{cm}^{-2}$ \footnote{See \url{https://heasarc.gsfc.nasa.gov/cgi-bin/Tools/w3nh/w3nh.pl}}. 

\begin{table*}
\caption{The MeerKAT radio flux densities (1.28 GHz) and luminosities at 5GHz, along with Swift and NICER X-ray flux and luminosities matches from the 1-10 keV energy band. The upper limits are 3$\sigma$ and the uncertainties in the table are 1$\sigma$. The luminosities are based on distance estimates of 3.5\,kpc, 10.4\,kpc and 10\,kpc for SAX J1808, MAXI J0911 and XTE J1701, respectively. The Swift and NICER radio:X-ray matches are indicated as S$^{1,2,3}$ and N$_{1,2,3,4,5,6}$ respectively, next to the MJDs. }
\begin{tabular}{l c c c c c c}\hline
MJD & F$_{\rm{r}}$ [$\rm{mJy/beam}$] & F$_{\rm{x\textit{NICER}}}\times 10^{-10}$  &F$_{\rm{x\textit{Swift}}}\times 10^{-10}$ 	&L$_{\rm{r}}\times 10^{28}$	&	L$_{\rm{x\textit{NICER}}}\times 10^{36}$  &L$_{\rm{x\textit{Swift}}}\times 10^{36}$ 	\\
 & & [erg/s/cm$^2$] &[erg/s/cm$^2$] & [erg/s] &[erg/s]&[erg/s] \\\hline
 SAX J1808 & & & & & \\\hline
 58695.02$_{N1}$ & & 0.0040 $\pm$ 0.0002 &  & & 0.00059 $\pm$ 0.00003\\
 58695.74$_{N1}$ & $<$0.075 & &  & $<$0.56 & \\
 58699.27$_{N2}$ & & 0.0060 $\pm$ 0.0002 &  & & 0.00088 $\pm$ 0.00003 \\\\
 58699.82$_{N2}$ & $<$0.12 & &  & $<$0.86  & \\
 58705.01$_{N3}$ & & 6.20 $\pm$ 0.02 &  & & 0.910 $\pm$ 0.003 \\\\
 58705.81$^{S1}_{N3}$ & 0.59	$\pm$	0.04 &	&  &	4.32	$\pm$	0.25 & 	\\\\
 58708.47$^{S1}$ &  & &7.27 $\pm$ 0.16 &  &	&	1.065$^{+0.023}_{-0.024}$		\\\\
 58710.40 & & & 6.33 $\pm$ 0.16 &	& &	0.930	$\pm$	0.023		\\
 58711.05$_{N4}$ & & 4.70 $\pm$ 0.01 &  & & 0.700 $\pm$ 0.002 \\\\
 58711.95$^{S2}_{N4}$ & 0.36	$\pm$	0.02 & & 	&	2.67	$\pm$	0.15 & 	\\\\
 58712.65$^{S2}$ &  & & 3.98 $\pm$ 0.14 & & 	&	0.58 $\pm$	0.02	\\
 58718.70$_{N5}$ & 0.11	$\pm$	0.02	& &	 &	0.83	$\pm$	0.15 & 	\\
 58719.00$_{N5}$ & & 0.920$\pm$0.005 &  & & 0.1300$\pm$0.0007 \\\\
 58726.79$^{S3}_{N6}$ & 0.16	$\pm$	0.02	& & 	&	1.20	$\pm$	0.13 &	\\\\
 58726.98$_{N6}$ & & 0.80 $\pm$ 0.02 &  & & 0.20$\pm$ 0.02 \\\\
 58728.45$^{S3}$ &  & & 1.390$^{+0.063}_{-0.066}$& & &		0.2030$^{+0.0092}_{-0.0096}$	\\\\
 58740.81	&	& &	0.937$^{+0.078}_{-0.065}$	&		& &	0.137$^{+0.012}_{-0.010}$\\
 58741.79 & & &  & & \\
\hline
MAXI J0911 \\\hline
59403.25 & $<$0.055 & & $<$2.32 &   $<$3.57 & & $<$3 \\\hline
XTE J1701  \\\hline
59421.75 & $<$0.072 &  &$<$2094 &   $<$4.29 & & $<$2500 \\\hline
    \end{tabular}
    \label{tab:lums}
\end{table*}

\section{Results}

\subsection{SAX J1808}
In the X-rays, our \textit{NICER} results reveals the outburst of SAX J1808 to have a main outburst and subsequent reflaring (top panel of Figure~\ref{fig:SAX_flux}) within the 1 -- 10.0\,keV energy band. The main outburst rises around 6 August 2019 (MJD = 58701) and the reflaring around 27 August 2019 (MJD = 58722). The 5 detections we record here from \textit{Swift} only display the decline of the main outburst with a peak flux at (7.27$\pm$0.16) $\times 10^{-10}$\,erg/s/cm$^2$ on the 12 August 2019 (MJD = 58708.47, second panel of Figure~\ref{fig:SAX_flux}). While our \textit{NICER} results reveal a peak flux (6.20$\pm$0.018)$\times 10^{-10}$\,erg/s/cm$^2$ (247 counts/s, MJD = 58705.01 in the first panel of Figure~\ref{fig:SAX_flux}), similarly \cite{2019ApJ...885L...1B} record a peak bolometric flux of (4.7$\pm$0.5)$\times 10^{-10}$\,erg/s/cm$^2$ within the 0.01 -- 100\,keV energy band.
SAX J1808 was undetected in the first two epochs of the MeerKAT monitoring. These non-detections have upper limits of 0.075\,mJy/beam and 0.12\,mJy/beam at 3$\sigma$, respectively (indicated with upside-down triangles in the third panel of Figure~\ref{fig:SAX_flux}). The lightcurve was constructed using the flux measurements from {\textsc{pybdsf}} such that the radio flare is seen to peak at 0.59$\pm$0.034\,mJy/beam. We observe a decline in the radio lightcurve between 10 -- 23 August (MJD = 58705 -- 58718) but the rise of a reflare event is seen between the 5$^{\rm th}$ and 6$^{\rm th}$ epoch (23 -- 31 August, MJD = 58718 -- 58726) as an increase from 0.11 -- 0.16\,mJy/beam is measured. 
In addition to the LCO optical data, the source was observed with MeerLICHT, a prototype of the BlackGEM array \citep{2016SPIE.9906E..64B,2019NatAs...3.1160G} which provides simultaneous monitoring of the same area of the sky as MeerKAT. The data were processed automatically with the \textsc{BlackBOX/ZOGY} pipeline \citep{2021ascl.soft05011V}, which performs primary reductions that include bias subtraction, overscan corrections and flat-fielding, as well as astrometric and photometric calibrations. Daily averages of each band were determined by co-adding the images and ranged between 17.89$\pm$0.04 -- 19.38$\pm$0.06 magnitudes across all bands. The results from MeerLICHT, which we do not plot, confirm the results from the more comprehensive data from LCO displayed in the bottom panel of Figure~\ref{fig:SAX_flux}.
\noindent
\\\\
\subsection{MAXI J0911 and XTE J1701}
\begin{figure*}

    \centering
    \includegraphics[scale=0.5]{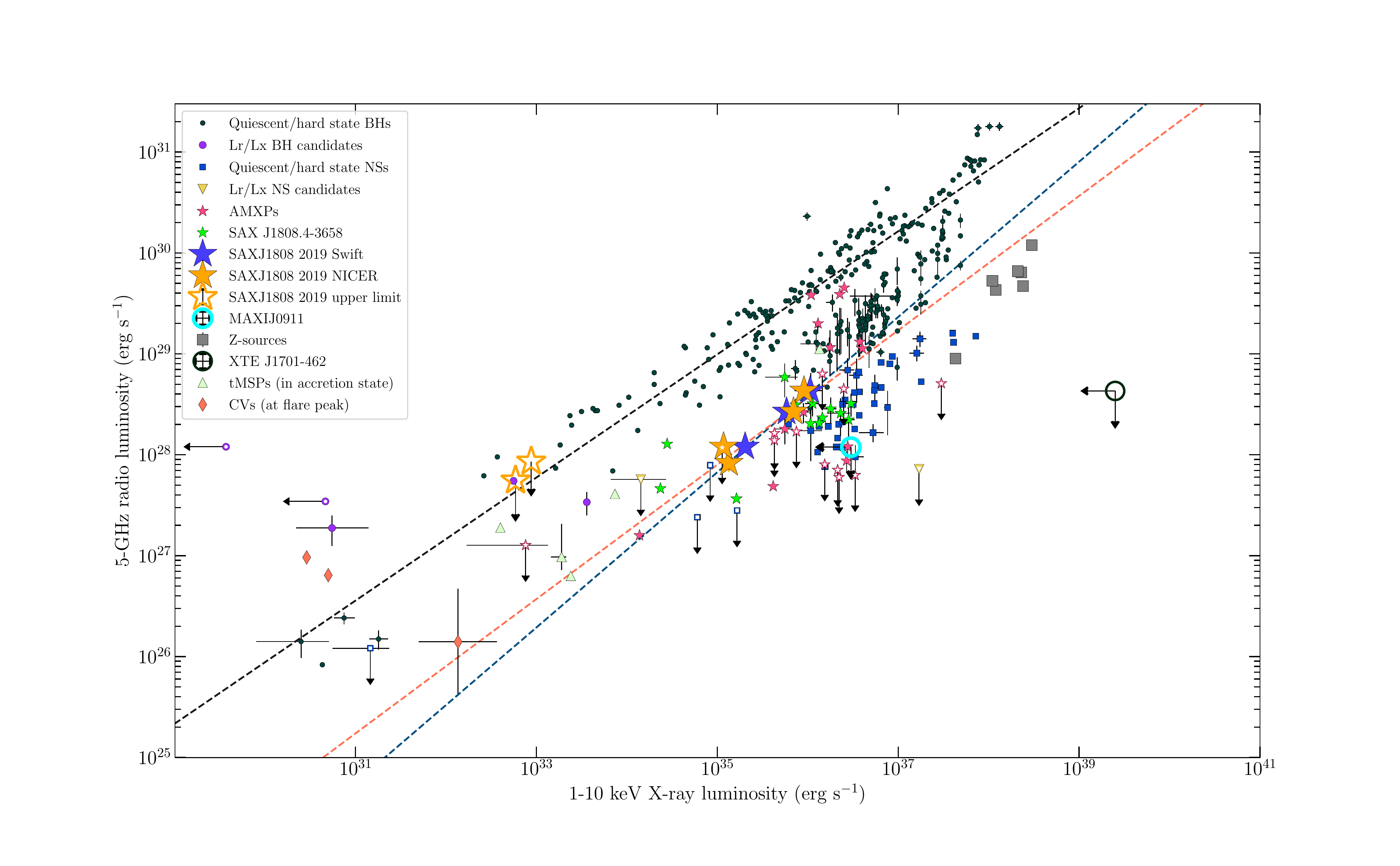}
    \caption{The radio:X-ray correlation for BH and NS sources labeled in terms of class \citep{2006MNRAS.366...79M,arash_bahramian_2018_1252036}, with previous SAX J1808 detections indicated with green stars and the results from this work with the blue and orange stars, with the dashed blue line indicating the slope of the \textit{Swift} matches and the orange dashed line the slope of the \textit{NICER} detection matches. The \textit{NICER} upper limit matches are the open orange stars. The BH population slope is indicated with the black dashed line and the upper limits from the MAXI J0911 and XTE J1701 observations are indicated with the cyan and black open circles respectively}
    \label{fig:SAX_radioxraylum}
\end{figure*}

In the radio, neither MAXI J0911 nor XTE J1701 were detected. The 3$\sigma$ upper limits presented in Table~\ref{tab:lums} represent that of the single radio:X-ray matches for each source. Assuming a flat spectral index the radio flux was converted to a frequency of 5 GHz such that the luminosity $\nu$L$_{\nu}$ is $5$L$_{5}$ (referred to as radio luminosity L$_{\rm R}$ in the paper). The luminosity upper limits are L$_{\rm R}$ $<$ $3.57\times 10^{28}$\,erg/s and L$_{\rm R}$ $<$ $4.29\times 10^{28}$\,erg/s taken on MJD 59421.65 and 59421.76 respectively (26 July 2021) for MAXI J0911 and XTE J1701 respectively. These luminosities were determined using distances 10.4\,kpc \citep{2016ATel.8914....1T} and 10\,kpc \citep{2007MNRAS.380L..25F} for MAXI J0911 and XTE J1701 respectively and the upper limits are plotted in cyan (MAXI J0911) and black open circles (XTE J1701) in Figure~\ref{fig:SAX_radioxraylum}.

\begin{table}
    \caption{The peak flux density of the extended sources B C D of XTE J1701 as indicated in Figure 5 of \protect\cite{2007MNRAS.380L..25F}, along with the spectral index between the MeerKAT (1.28\,GHz) 2021 observation and the ATCA (4.8, 8.6\,GHz) 2006 observations in \protect\cite{2007MNRAS.380L..25F}.}
    \begin{tabular}{llllll}\hline
         Source & F$_{1.28}$ & F$_{4.8}$ & F$_{8.6}$ & $\alpha$ & $\alpha$  \\
         & \big[$\rm{\frac{mJy}{beam}}$\big] & \big[$\rm{\frac{mJy}{beam}}$\big] & \big[$\rm{\frac{mJy}{beam}}$\big] & (4.8-8.6) & (1.28-4.8-8.6) \\\hline
         B & 22.23 & 2.5 & 0.9 & -1.7 & -1.7 \\
         C & 16.38 & 2.0 & 0.5 & -2.4 & -1.7 \\
         D & 16.55 & 1.5 & 0.3 & -2.7 & -1.9 \\\hline
    \end{tabular}
    
    \label{tab:XTE_fluxes}
\end{table}

\begin{figure}
    \centering
    \includegraphics[scale=0.49]{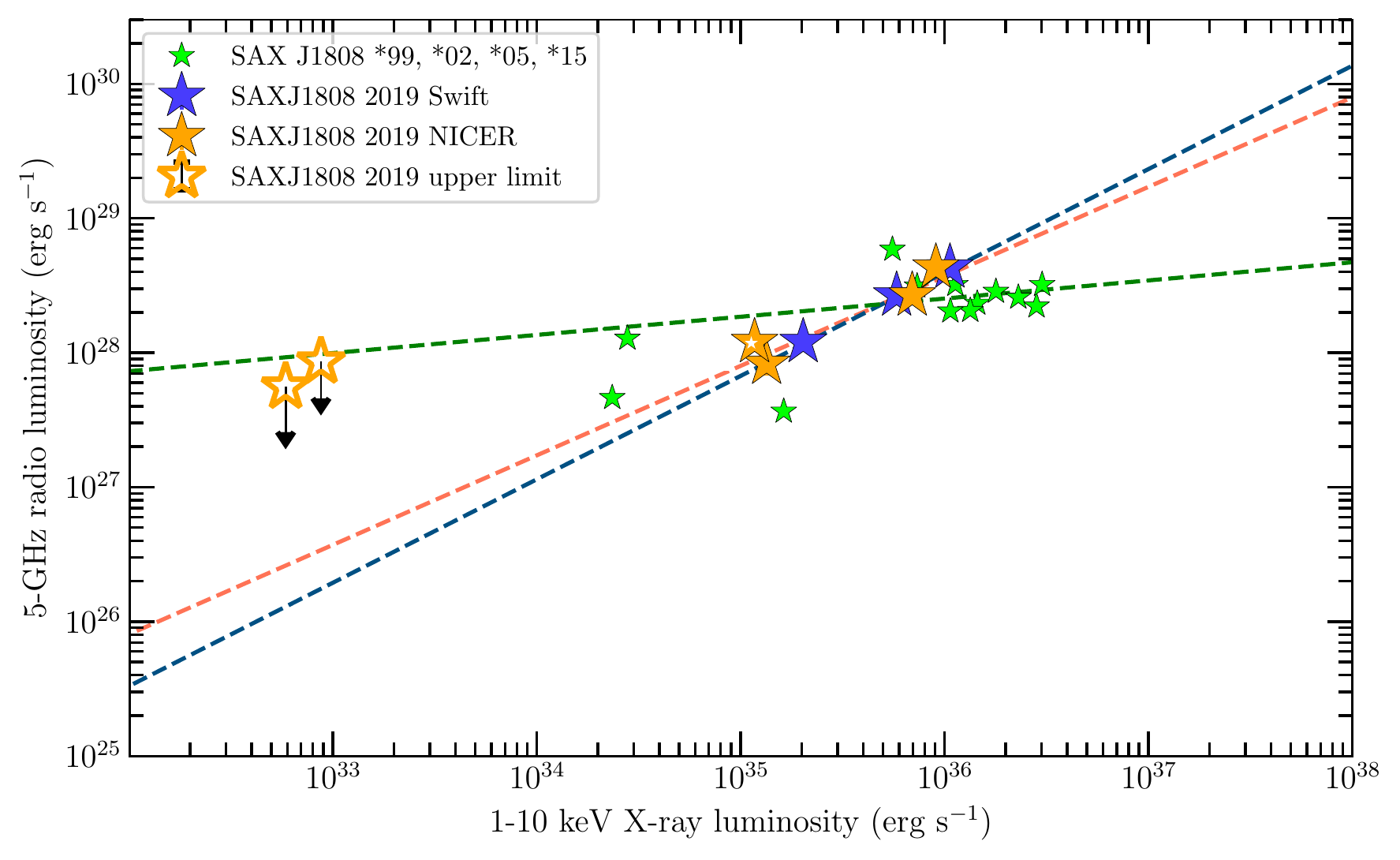}
    \caption{The radio:X-ray correlation slopes for the SAX J1808 outbursts; 2019 outburst (blue and orange), a group of four of the older outbursts reported (green). The combination of all older outburst points on the figure (1999, 2002, 2005, 2015) are fit to a slope of $\beta $=$ 0.33 \pm 0.23$ and indicated with the green dashed line. The blue and orange dashed lines represent the detection matches (excluding upper limits) with slopes $\beta $=$ 0.77 \pm 0.13$ and $\beta $=$ 0.67 \pm 0.33$ for the \textit{Swift} and \textit{NICER} 2019 outburst, respectively. The NICER radio upper limit matches are presented as open orange stars.}
    \label{fig:SAX_radioxraylumfit}
\end{figure}
\section{Discussion}
In this section we discuss the 2019 outburst of SAX J1808 and the placement of the sources (in this work) on the radio:X-ray correlation. We only provide single upper limits of MAXI J0911 and XTE J1701, but the 2019 outburst results significantly extend SAX J1808's X-ray luminosity range on the plane.
\subsection{SAX J1808 2019 Outburst}
Our results show that the X-ray peak follows the radio peak where maximum flux density is recorded as 0.59$\pm$0.034\,mJy/beam, indicating jet action from 10 August 2019 (MJD 58705). The optical reports of the 2019 outburst recorded (see \citealt{2020MNRAS.498.3429G,2020ApJ...905...87B}) observations from Las Cumbres Observatory (LCO) and the Southern African Large Telescope (SALT). \cite{2020MNRAS.498.3429G} caught the rise of the outburst with a peak optical flux $>$ 1.2\,mJy around the 10 August and we present the \cite{2020ApJ...905...87B} lightcurve in bottom panel of Figure~\ref{fig:SAX_flux} where we note two rise and dips occurrences (around 22 July -- 3 August 2019 corresponding to MJDs 58686 -- 58698) before the peak of the main outburst is seen around 11 August 2019 (MJD 58706). The reflaring event started around the 24 August 2019 and while we only see the start of this reflare in our radio results (third panel Figure~\ref{fig:SAX_flux}), it shows the jets are still present as suggested in \cite{2020ApJ...905...87B}. In \cite{2020MNRAS.498.3429G} the X-ray peak was detected on the 14 August 2019 (MJD 58709). These reports together with our results suggest the optical peaked first and the X-ray and radio outbursts peaked later. This result may be explained by an outside-in outburst (see \citealt{1999MNRAS.310..398I}) starting from the outer edges of the disk. The early optical activity in the 2019 outburst is discussed in more detail in \cite{2020MNRAS.498.3429G}, proposing scenarios that stem from mass transfer variations and geometrical effects of the disk. The cadence of our radio observations was not high enough to quantify the outside-in outburst mechanism. 

\subsection{The radio:X-ray correlation of SAX J1808}
Figure~\ref{fig:SAX_flux} illustrates the evolution of the radio (MeerKAT) and X-ray (\textit{NICER}, \textit{Swift}) outburst in the first three panels. A flat spectral index is assumed and the radio luminosity is determined similarly to MAXI J0911 and XTE J1701 mentioned in Section 3.2; in the case of SAX J1808 the luminosities are based on a distance of 3.5\,kpc \citep{2006ApJ...652..559G}. Based on the 10 quasi-simultaneous \textit{Swift} observations, the source was only detected in 6 epochs. After inspection, 3 epochs were selected which were within $\Delta$t $\sim2$\,days of 3 of the radio observations (see matches S$^{\rm{1,2,3}}$ in Table~\ref{tab:lums}). The radio and X-ray luminosities from the selected MJDs, were used to determine SAX J1808's position on the radio:X-ray correlation plot. Since the \textit{NICER} observations are taken at higher cadence, we were able to match the 6 radio observations within $\Delta$t $\sim1$\,day (see matches N$_{\rm{1,2,3,4,5,6}}$ in Table~\ref{tab:lums}). The radio:X-ray correlation is shown in Figure~\ref{fig:SAX_radioxraylum} where the blue stars indicate the 2019 outburst results based on the \textit{Swift} matches and the orange stars are the \textit{NICER} matches for the 6 radio observations. The previous L$_{\rm R}$L$_{\rm X}$ measurements of the source are shown with the green stars. In this figure SAX J1808 lies among the AMXP class of sources which are indicated with the pink stars. This is an expected result for the source since it is the cornerstone of the AMXP class, and we note that these results have less scatter than the previous outbursts. The upper limits of the \textit{NICER} matches extend to lower radio and X-ray luminosities, revealing how extensive the luminosity range of the source can get on the correlation plane. The {\textsc{python}} module, {\textsc{linear regression}}, in the {\textsc{sklearn}} library was used to determine a best fit of the SAX J1808 2019 outburst, the data points were fed into the model and as a result produced slopes $\beta $=$ 0.77 \pm 0.13$ and $\beta $=$ 0.67 \pm 0.33$ for the detection matches (excluding upper limits) of \textit{Swift} and \textit{NICER}, respectively. We also fit all the \textit{NICER} matches (including the upper limits) and the best fit slope was $\beta $=$ 0.35 \pm 0.38$. The older outbursts results produced a fit of ($\beta $=$ 0.33 \pm 0.23$), the combination of complete SAX J1808 points ($\beta $=$ 0.34 \pm 0.20$) and the complete AMXP class ($\beta $=$ 0.38 \pm 0.23$) only included the set of \textit{Swift} matches (Figure~\ref{fig:SAX_radioxraylumfit}). These less steep slopes are suspected to be due to the increase in scatter of the overall sample of points of the source. The slope which includes the low luminosity upper limits from \textit{NICER} is small too but with larger error, in contrast a recent report on intermittent AMXP Aql X-1 at low X-ray luminosities finds a slope of $\beta $=$ 1.17^{+0.30}_{-0.21}$ where detection and upper limits of multiple outbursts are included for the fit (see more details in \citealt{2020MNRAS.492.2858G}). \cite{2020MNRAS.492.2858G} also fit detections only, these slopes (based on different hardness ratios) ranged between $\beta \sim$ 0.73 -- 0.96. This indicates that a single NS-XRB can have a varying slope depending on the outburst and the detections included in the fit.  Furthermore, the radio:X-ray correlation for BHXBs and NS-XRBs from \cite{2011IAUS..275..255C}  reports the radiatively inefficient and efficient branches have slopes $\beta \sim 0.7$ and $\beta \sim 1.4 $ for the BHXBs and NS-XRBs respectively. Similarly seen in \cite{2003MNRAS.342L..67M}, while some NS-XRB systems follow this trend, the slopes of SAX J1808 in this work match more closely to that of the BHXBs. This is a result that has been observed before for NS-XRBs. \cite{2017MNRAS.470..324T} demonstrated that the slope of AMXP IGR J00291+5934 is 0.77 $\pm$ 0.18, hence it does appear the correlation slopes of some NSs behave differently. 

\subsection{MAXI J0911-655}
MAXI J0911, which resides in the globular cluster NGC 2808, was first detected with MAXI/GSC's nova alert system in February 2016 \citep{2011PASJ...63S.623M, 2016ATel.8872....1S}. The X-ray activity from the source was detected about a week later with \textit{Swift}/BAT \citep{2017A&A...598A..34S}. Later detections of X-ray activity prompted simultaneous radio follow-up with ATCA \citep{2016ATel.8914....1T}, yielding an upper limit L$_{\rm R}$ $<$ $4.5\times 10^{28}$\,erg/s at 5 GHz. Assuming a flat spectrum and distance 10.4\,kpc, we derive a 3$\sigma$ upper limit L$_{\rm R}$ $<$ $3.57\times 10^{28}$\,erg/s based on MeerKAT observations taken in July 2021 (see cyan open circle in Figure~\ref{fig:SAX_radioxraylum}) after weak hard X-rays were detected in \cite{2021ATel14767....1N}.

\subsection{XTE J1701}
XTE J1701 is an important source in the study of NS-XRBs as it exhibited behaviour associated with both 'Z' and 'Atoll'-type binaries, establishing that these classifications depend upon accretion state \citep{2009ApJ...696.1257L}. \cite{2007MNRAS.380L..25F} reported a small number of radio detections of XTE J1701, which were broadly consistent with those reported for other Z-sources. They also reported an extended structure (labelled sources B C and D in Figure 5 of \cite{2007MNRAS.380L..25F}) approximately three arcmin to the south which they speculated could be a large-scale jet associated with the outburst of the binary. This structure is visible in our MeerKAT image at flux densities consistent with being constant over $\sim 15$ years (see luminosity upper limits at 10\,kpc \citep{2007MNRAS.380L..25F} in Table~\ref{tab:lums} and spectral indexes Table~\ref{tab:XTE_fluxes}, while the source is placed as the black open circle on the radio:X-ray plane in Figure~\ref{fig:SAX_radioxraylum}). We therefore rule out any association with a transient jet associated with the binary, and attribute these sources to background extragalactic sources instead. The limit on the core radio luminosity from XTE J1701 is not particularly constraining given the quiescent state during the July 2021 observation.

\section{Conclusion}
NS-XRBs are an interesting class to study the disk/jet coupling since the radio:X-ray correlation tends to vary considerably in different sources and accretion regimes. We report here on the 2019 outburst of SAX J1808 as well as follow up observations of MAXI J0911 and XTE J1701 carried out as part of the ThunderKAT Large Survey Programme on MeerKAT \citep{2016mks..confE..13F}.
SAX J1808 is reported in the radio, X-ray and the optical from MeerKAT, \textit{Swift}, \textit{NICER} and LCO respectively. The outburst lasted from July to early October 2019, including a main outburst event and a reflare event which started late August to early September 2019. These results suggest the peak of the main outburst is seen in the optical before the X-ray and radio which may imply an outside-in outburst \citep{1999MNRAS.310..398I}, however, our radio observations were not sampled at a high enough cadence to suggest the mechanism occurred in this system. Furthermore, the reflaring event was associated with a radio re-brightening, indicating a strengthening of the jet in this phase \citep{2020ApJ...905...87B}. The radio:X-ray correlation of the source in this work (\textit{Swift} and \textit{NICER} detection matches have a slope of $\beta \sim 0.77 \pm 0.13$ and $\beta \sim 0.67 \pm 0.33$ respectively) suggests a similar radiative efficiency compared to the BHXB standard slope of $\beta \sim 0.6$ \citep{2000A&A...359..251C,2003A&A...400.1007C,2013MNRAS.428.2500C,2021MNRAS.505L..58C}. Finally, the upper limits of MAXI J0911 and XTE J1701 from both MeerKAT and \textit{Swift} place the sources in their respective classes (AMXP the former, Z/Atoll the latter). And while these are non-detections the results encourage continued monitoring of these sources and similar NS-XRBs to improve our understanding and reveal the true nature of the disk and jet coupling in XRBs.

\section*{Acknowledgements}
KVSG, IMM, PAW and PJG acknowledge support from the University of Cape Town and the National Research Foundation. KVSG would like to thank Dave Russell for making the LCO data published in \cite{2020ApJ...905...87B} available for the improvement of this paper. KVSG acknowledges Lauren Rhodes for the advice during radio data processing. PJG is supported by NRF SARChI grant 111692. SEM acknowledges financial contribution from the agreement ASI-INAF n.2017-14-H.0, and PRIN-INAF 2019 n.15. DRAW was supported by the Oxford Centre for Astrophysical Surveys, which is funded through generous support from the Hintze Family Charitable Foundation. JvdE is supported by a Lee Hysan Junior Research Fellowship awarded by St. Hilda's College, Oxford. IH acknowledges support from the UK Science and Technology Facilities Council [ST/N000919/1], and from the South African Radio Astronomy Observatory. VAM acknowledges support from the National Research Foundation grant 119466.

\noindent
The MeerKAT telescope is operated by the South African Radio Astronomy Observatory, which is a facility of the National Research Foundation, an agency of the Department of Science and Innovation. We acknowledge the use of public data from the \textit{Swift}/BAT/XRT data archive.
We acknowledge the use of the ilifu cloud computing facility – www.ilifu.ac.za, a partnership between the University of Cape Town, the University of the Western Cape, the University of Stellenbosch, Sol Plaatje University, the Cape Peninsula University of Technology and the South African Radio Astronomy Observatory. The Ilifu facility is supported by contributions from the Inter-University Institute for Data Intensive Astronomy (IDIA – a partnership between the University of Cape Town, the University of Pretoria, the University of the Western Cape and the South African Radio astronomy Observatory), the Computational Biology division at UCT and the Data Intensive Research Initiative of South Africa (DIRISA).
\section*{Data Availability}

The data underlying this article will be shared on reasonable request to the corresponding author.



\bibliographystyle{mnras}

\bibliography{references}



\twocolumn
\appendix

\section{Some extra material}



\bsp	
\label{lastpage}
\end{document}